\documentclass[10pt,a4paper]{article}
\usepackage{amsmath,amssymb,amsthm,eucal,cite}
\usepackage[top=3cm,right=3cm,left=2.5cm,bottom=3cm]{geometry}
\usepackage{tikz}
\usetikzlibrary{calc,matrix,patterns}

\title{Perturbative QFT: off-shell fields, deformation quantization and causal perturbation theory%
\footnote{This short review is commissioned by the Encyclopedia of Mathematical Physics,
edited by M. Bojowald and R.J. Szabo, to be published by Elsevier.}}

\author{Michael D\"utsch%
\footnote{Institute f\"ur Theoretische Physik, Universit\"at G\"ottingen, 37077 G\"ottingen, Germany;\newline  
e-mail: michael.duetsch3@gmail.com}}

\date{}

\newcommand{\be}{\begin{equation}}   
\newcommand{\ee}{\end{equation}} 

\newcommand{\al}{\alpha}          %% short for \alpha
\newcommand{\bt}{\beta}           %% short for \beta
\newcommand{\Dl}{\Delta}          %% short for \Delta
\newcommand{\dl}{\delta}          %% short for \delta
\newcommand{\ka}{\kappa}          %% short for \kappa
\newcommand{\La}{\Lambda}         %% short for \Lambda
\newcommand{\la}{\lambda}         %% short for \lambda
\newcommand{\om}{\omega}          %% short for \omega
\renewcommand{\th}{\theta}        %% short for \theta
\newcommand{\vf}{\varphi}         %% short for \varphi

\newcommand{\bC}{\mathbb{C}}      %% complex numbers
\newcommand{\bE}{\mathbb{E}}      %% Euler operator
\newcommand{\bM}{\mathbb{M}}      %% Minkowski space
\newcommand{\bN}{\mathbb{N}}      %% natural numbers
\newcommand{\bR}{\mathbb{R}}      %% real numbers
\newcommand{\gF}{\mathfrak{F}}    %% Fock space
\newcommand{\gS}{\mathfrak{S}}    %% symmetrization

\newcommand{\nS}{\mathbf{S}}      %% S in `S-matrix'

\newcommand{\sA}{\mathcal{A}}     %% an algebra
\newcommand{\sC}{\mathcal{C}}     %% smooth functions
\newcommand{\sD}{\mathcal{D}}     %% space of test functions
\newcommand{\sF}{\mathcal{F}}     %% set of observables
\newcommand{\sG}{\mathcal{G}}     %% set of coupling functions
\newcommand{\sL}{\mathcal{L}}     %% a Lagrangian, Lorentz group
\newcommand{\sO}{\mathcal{O}}     %% an open region
\newcommand{\sP}{\mathcal{P}}     %% set of polynomials
\newcommand{\sR}{\mathcal{R}}     %% renormalization group
\newcommand{\bal}{\mathrm{bal}}   %% balanced fields
\newcommand{\homog}{\mathrm{hom}} %% homogeneous polynomials
\newcommand{\intr}{\mathrm{int}}  %% interacting things
\newcommand{\loc}{\mathrm{loc}}   %% local fields
\newcommand{\MS}{\mathrm{MS}}	  %% minimal subtraction
\newcommand{\op}{\mathrm{op}}     %% operator-valued fields
\newcommand{\pp}{\mathrm{pp}}     %% principal part
\newcommand{\rp}{\mathrm{rp}}     %% regular part
\newcommand{\red}{\mathrm{red}}   %% `reduced' quantity
\newcommand{\ret}{\mathrm{ret}}   %% retarded fields or propagators
\newcommand{\bigveestar}{\bigvee\nolimits_{\!\star}} %% algebra maker
\newcommand{\del}{\partial}       %% short for \partial
\newcommand{\downto}{\downarrow}  %% short for \downarrow
\newcommand{\longto}{\longrightarrow}
\newcommand{\less}{\setminus}     %% set difference
\newcommand{\ovl}{\overline}      %% short for \overline
\newcommand{\ox}{\otimes}         %% tensor product
\newcommand{\oxyox}{\otimes\cdots\otimes} %% repeated tensor product
\newcommand{\unl}{\underline}     %% short for \underline
\newcommand{\up}{{\mathord{\uparrow}}} %% forward time
\newcommand{\wh}{\widehat}        %% short for \widehat
\newcommand{\x}{\times}           %% cartesian product or cross
\renewcommand{\.}{\cdot}          %% dot product
\renewcommand{\:}{\colon}         %% colon in  f: A -> B

\DeclareMathOperator{\Id}{Id}      %% identity operator
\DeclareMathOperator{\sd}{sd}      %% scaling degree
\DeclareMathOperator{\supp}{supp}  %% support
\DeclareMathOperator{\WF}{WF}      %% wave front set

\newcommand{\ldbrack}{[\mskip-2.5mu[} %% double brackets [[
\newcommand{\rdbrack}{]\mskip-2.5mu]} %% double brackets ]]

\newcommand{\ddto}[1]{\frac{d}{d#1}\biggr|_{#1=0}} %% derivative at 0
\newcommand{\pw}[1]{\ldbrack#1\rdbrack} %% power series ring
\newcommand{\set}[1]{\{\,#1\,\}}    %% set notation
\newcommand{\Set}[1]{\biggl\{#1\biggr\}} %% bigger set notation
\newcommand{\word}[1]{\quad\mbox{#1}\quad} %% well-spaced word(s)
\def\wick:#1:{\mathopen:#1\mathclose:} %% normally ordered expression

\newcommand{\fd}[2]{\frac{\dl#1}{\dl\vf(#2)}} %% \dl f/\dl\vf(x)
\def\duo<#1,#2>{\langle#1,#2\rangle} %% pairing of distributions

\numberwithin{equation}{section}

\theoremstyle{plain}
\newtheorem{thm}{Theorem}[section]  %% Theorem 1.1
\newtheorem{prop}[thm]{Proposition} %% Proposition 1.2
\theoremstyle{definition}
\newtheorem{defn}[thm]{Definition}  %% Definition 1.6

\newtheoremstyle{example}%          %% examples in \small size
   {\topsep}{\topsep}{\small}{0pt}%
   {\bfseries}{.}{ }{}
\theoremstyle{example}
\theoremstyle{remark}
\DeclareRobustCommand{\qned}{\ifmmode
  \else \leavevmode\unskip\penalty9999 \hbox{}\nobreak\hfill \fi
  \quad\hbox{\qnedsymbol}}
\newcommand{\qnedsymbol}{$\boxminus$} %% Non-proofs end with `\qned'

\makeatletter
\renewcommand{\section}{\@startsection{section}{1}{\z@}%
                        {-3.5ex \@plus -1ex \@minus -.2ex}%
                        {2.3ex \@plus.2ex}%
                        {\normalfont\large\bfseries}}
\renewcommand{\subsection}{\@startsection{subsection}{2}{\z@}%
                        {-3.25ex \@plus -1ex \@minus -.2ex}%
                        {1.5ex \@plus .2ex}%
                        {\normalfont\normalsize\bfseries}}
\renewcommand{\subsubsection}{\@startsection{subsubsection}{3}{\z@}%
                        {-3.25ex \@plus -1ex \@minus -.2ex}%
                        {1.5ex \@plus .2ex}%
                        {\normalfont\normalsize\itshape}}
\renewcommand{\@dotsep}{200} %% suppress dots in Contents
\makeatother

\begin{document}

\maketitle

\begin{abstract}
Perturbative QFT is developed in terms of off-shell fields (that is, functionals on the configuration space not restricted by any field equation),
and by quantizing the (underlying) free theory by an $\hbar$-dependent deformation of the classical product (i.e., the pointwise
product of functionals). The time-ordered product of local fields is defined axiomatically, and constructed by induction on the number of factors using Stora's version of the Epstein--Glaser construction; in particular, the interaction is adiabatically switched off. The set of 
solutions of these axioms can be understood as the orbit of the Stückelberg--Petermann renormalization group  when acting 
on a particular solution. Interacting fields are defined in terms of the time-ordered product by Bogoliubov's formula; they satisfy 
the following, physically desired properties: causality, spacelike commutativity, (off-shell) field equation and existence of the classical 
limit.  Local, algebraic properties of the observables can be obtained without performing the adiabatic limit (i.e., the limit removing the 
adiabatic switching of the interaction). 
\end{abstract}

\section{Introduction}

Causal perturbation theory is based on ideas of Stückelberg and  Bogoliubov \cite{BogoliubovS59} which were
rigorously worked out in the seminal paper of Epstein and Glaser (EG) \cite{EpsteinG73}.
It was further developed mainly by R.~Stora (e.g., \cite{PopineauS16,Stora08,NikolovST14}) and the groups of G.~Scharf \cite{Scharf}
and K.~Fredenhagen \cite{BF00,BrunettiDF09,DuetschF99-04,DuetschFKR14}. Causal perturbation theory is a rigorous 
perturbative approach to Quantum Field Theory (QFT) -- winning by its \emph{conceptual clarity}. 
The latter relies on the following properties:
\begin{itemize}
\item The time-ordered product ($T$-product), which is the main building stone of a perturbative QFT, 
is defined by \emph{axioms}, the most important being Causality.
\item The interaction is adiabatically switched off. By this, the infrared (IR) problem is 
separated from the ultraviolet (UV) problem. The adiabatic limit (i.e., the limit which removes this unphysical 
switching of the interaction -- this is the IR problem) is performed only at the end of the construction; typically it exists only for
observable quantities, e.g., inclusive cross sections, and not for the individual $S$-matrix elements. However, local, algebraic properties
of the observables can be obtained without performing the adiabatic limit, see Sect.~\ref{sec:ad-lim}. After performing 
the adiabatic limit, the results agree with what comes out from more conventional versions of perturbative QFT,
e.g., BPHZ-renormalization \cite{BogoliubovP57,Hepp66,Zimmermann69} or dimensional regularization.
\item The $T$-product $T=(T_n)_{n\in\bN}$ is constructed in position space, 
by induction on the number $n$ of factors. Due to this, renormalization (i.e., the UV-problem) is the mathematically 
well-defined problem of extending inductively known distributions from $\sD'(M^n\less\Dl_n)$ to $\sD'(M^n)$,
where $M$ is the space-time manifold and $\Dl_n$ is the thin diagonal in $M^n$, see \eqref{eq:Dln}. 
As long as one does not consider the adiabatic limit, in each step, all quantities are mathematically well-defined.
\item The observables are constructed as \emph{formal} power series in the coupling constant and in $\hbar$ 
-- questions concerning the convergence of this series are not touched. 
\item The EG-construction yields \emph{all} solutions of the axioms. By the Main Theorem (Thm.~\ref{th:main-thm-renorm}) 
the set of solutions is the orbit of the Stückelberg--Petermann renormalization group (Def.~\ref{df:SP-RG}) when acting 
on a particular solution (any solution may be chosen as starting point).
\end{itemize}

Further advantages of the EG-construction of the $T$-product are:
\begin{itemize}
\item Since it proceeds in position space, it is well suited for perturbative QFT 
on a globally hyperbolic, curved space-time manifold $M$ \cite{BF00, HollandsW01-05} -- see the next article in this encyclopedia.
For simplicity, in this article, we choose $M$ to be the $d$-dimensional Minkowski space. EG-renormalization has been 
worked out also in Euclidean space \cite{Keller09}.
\item Troubles with overlapping divergences do not appear, due to the inductive procedure in the construction of the $T$-product.
(Note that, speaking diagrammatically, the induction is w.r.t.~the number of vertices -- which contrasts with the inductive procedure 
in BPHZ-renormalization \cite{BogoliubovP57,Hepp66,Zimmermann69}.)
\item It applies also to nonrenormalizable interactions, e.g.~perturbative quantum gravity (see e.g.~\cite{Scharf,BrunettiFR16-23}):  
In each order in the coupling constant it yields a well-defined result. 
\end{itemize}

In most formulations of perturbative QFT (also in the work of EG \cite{EpsteinG73}) the free quantum field is a Fock space
operator. That is, it is an ``on-shell field'', since it obeys the free field equation. In this article, (classical and quantum) fields
are functionals on the classical configuration space, which is $C^\infty(M,\bR)$ in case of a real scalar field.
That is, our fields are ``off-shell'', since they are not restricted by any field equation. The algebra of classical fields is given by the
pointwise product of functionals. Quantization of the free theory is obtained by \emph{deformation} of this product, where
the propagator of the resulting star product (i.e., the two-point function) contains the information 
that we are dealing with the free theory (see \eqref{df:star-product}). Working with off-shell fields
is more flexible than the Fock space formalism; this is advantagous for various purposes -- see \cite[Preface]{D19}.

The main reference is the book \cite{D19}, which for the most parts relies on \cite{BF00,BrunettiDF09,DuetschF99-04,DuetschFKR14}. 
In the following, solely references differing from the just mentioned ones are given.

\section{Fields as functionals on the configuration space}

\paragraph{Space of fields.}
For brevity we study the model of one real scalar field with mass $m\geq 0$. 
For the pertinent configuration space  we choose $\sC:=\sC^\infty(M,\bR)$.
The partial derivative $\del^a$ (where%
\footnote{We use the French convention that $0\in\bN$.} 
$a\in\bN^d$) of the basic field $\vf(x)$ is the functional%
\footnote{The application of a functional (i.e., field) to a smooth function (i.e., configuration) is denoted by the $[\,\cdot\,]$-bracket.} 
\be\label{eq:dvf-functional}
\del^a\vf(x)\,\: \begin{cases}\sC \longto\bR \\
h \longmapsto \del^a\vf(x)[h]=\del^a h(x)\ .\end{cases}
\ee
 For simplicity, we
study only fields which are polynomials in the basic field~$\vf$ (and its derivatives). 
\begin{defn}
\label{df:fields}
The \emph{space of (classical and quantum) fields} $\sF$ is defined as
the vector space of functionals $F \equiv F(\vf) \: \sC \longto \bC$ of the form
\begin{align}
F(\vf) = f_0+\sum_{n=1}^N \int d^dx_1 \cdots d^dx_n \,\,\vf(x_1) \cdots \vf(x_n)\,
f_n(x_1,\dots,x_n)
=: \sum_{n=0}^N\langle f_n,\vf^{\ox n}\rangle
\label{eq:fields} 
\end{align}
with  $N < \infty$, where $F(\vf)[h]:=F(h)= \sum_{n}\langle f_n,h^{\ox n}\rangle$. Here $f_0 \in \bC$ is a
constant and, for $n \geq 1$, $f_n$ is a distribution (i.e., $f_n\in\sD'(M^n,\bC)$) 
with compact support. In addition, each $f_n$ is required to satisfy the
wave front set property:
\be
\WF(f_n) \subseteq \set{ (x_1,\dots,x_n; k_1,\dots,k_n) \,\big\vert\,
(k_1,\dots,k_n) \notin \ovl{V}_+^{\,\x n} \cup \ovl{V}_-^{\,\x n}}\ ,
\label{eq:wavy-front} 
\ee
where $V_\pm$ denotes the forward/backward light cone.
Convergence in $\sF$ is understood in the pointwise sense:
$\lim_{n\to\infty}F_n=F$ iff $\lim_{n\to\infty}F_n[h]=F[h]\,,\,\,\forall h\in\sC$.
\end{defn}

The purpose of the wave front set condition is to
ensure the existence of pointwise products of distributions which appear
in our definition of the Poisson bracket \eqref{eq:Poisson-br} and, more generally, of
the star product (see Sect.~\ref{sec:defo-quant}).

An important example is given by
$$
f_n(x_1,\ldots,x_n) := (-1)^{\sum_j|a_j|}\int dx\,\, g(x) \,\del^{a_1}\dl(x_1 - x) \cdots
\del^{a_n}\dl(x_n - x)\ ,\quad g\in \sD(M,\bC)\ ,
$$
and $f_k = 0$ for $k \neq n$, that is,
\be\label{eq:loc-func}
F(\vf) = \int dx\, g(x) \,\del^{a_1}\vf(x)
\cdots \del^{a_n}\vf(x)\in\sF\ .
\ee

The \emph{support} of~$F\in\sF$ is defined as
$$
\supp F := \supp \frac{\dl F}{\dl\vf(\.)}\ .
$$

\paragraph{Algebra of classical fields.}
Introducing the pointwise product 
\be\label{eq:class-product}
F\cdot G\equiv FG : h \longmapsto F[h]G[h]\in\sF ,
\ee
and the ``\mbox{$*$-operation}''
\be
F=\sum_{n=0}^N\langle f_n,\vf^{\ox n}\rangle\longmapsto F^*=\sum_{n=0}^N\langle \ovl{f_n},\vf^{\ox n}\rangle\in\sF
\ee
we obtain  a \emph{commutative \mbox{$*$-algebra}} -- this is the algebra of
classical fields.

\paragraph{Local fields.}
The example \eqref{eq:loc-func} is a local functional in the sense of the following definition:
\begin{defn}
\label{df:local-fields}
The space $\sF_\loc$ of \emph{local fields} is
following subspace of $\sF$: Let $\sP$ be the space of polynomials in the variables
$\{\del^a\vf\,|\,a \in \bN^d\}$ with real
coefficients (``field polynomials''); then
\begin{equation}
\sF_\loc := \Set{\sum_{i=1}^K A_i(g_i):=\sum_{i=1}^K \int dx\,\, A_i(x)\, g_i(x)
\,\Big\vert\, A_i \in \sP, \ g_i \in \sD(M,\bC), \ K < \infty}.
\label{eq:local-fields} 
\end{equation}
\end{defn}

Given $F = \sum_{i=1}^K A_i(g_i)\in\sF_\loc$,
the pairs $(A_i,g_i)_{i=1}^K$ are not uniquely determined by $F$, since
$\int dx\, \del_\mu\bigl(A(x)g(x)\bigr) = 0$ for any $A \in \sP$ and
$g \in \sD(M)$. This non-uniqueness can be removed in the following way:
\begin{prop}\label{prop:balanced}
There exists a subspace $\sP_\bal$ of $\sP$ (the space of ``\emph{balanced fields}'') with the following properties:
\begin{enumerate}
\item[\textup{(a)}]
Every $0\not= A \in \sP$ can \emph{uniquely} be written as a finite sum of
type
\be\label{eq:A-bal-simple}
A=\sum_{a\in\bN^d}\del^a B_a\  ,\word{where} B_a\in \sP_\bal
\word{and} B_a\vert_{\vf=0}=0\,\,\,\forall a\not= 0\ .
\ee
\item[\textup{(b)}]
For each $F \in \sF_\loc$, there exists a \emph{unique} $f\in\sD(M,\sP_\bal)$
(i.e., $f(x)=\sum_{k=1}^K g_k(x)\,B_k(x)$ with $g_k\in\sD(M,\bC),\,B_k\in\sP_\bal,\,K<\infty$)
such that
$$
F - F[0] = \int dx\, f(x) \word{and also} f(x)|_{\vf=0} = 0\quad\forall
x\in M\ .
$$
\end{enumerate}
\end{prop}
 
For example, $\vf\del^\mu\vf$ cannot be a balanced field, since
$\vf\del^\mu\vf=\del^\mu(\frac{1}2\,\vf^2)$.
In part~(b), $F[0] \in \bC$ must be excluded, since there are
infinitely many $\tilde f \in \sD(\bM,\bC)$ fulfilling
$F(0) = \int dx\, \tilde f(x)$.  It is an easy exercise to prove that part (a) implies part (b). 
Part (a) has been proved by giving an explicit construction of $\sP_\bal$.
%, see \cite{DuetschF99-04} or \cite[Chap.~1.4]{D19}. 

\paragraph{Poisson bracket.}
Let $-\Dl^\ret_m \in \sD'(\bR^d)$ be the retarded Green' function of the Klein--Gordon operator and 
let $\Dl_m(x):=\Dl_m^\ret(x)-\Dl_m^\ret(-x)$ be the commutator function.
\begin{defn} % 1.22
\label{df:Poisson-bracket}
The \emph{Poisson bracket} of the free theory
is the bilinear map $\sF \x \sF \longto \sF$ given by
\begin{equation}\label{eq:Poisson-br}
\{F, G\} := \int dx\,dy\,\, \fd{F}{x}\, \Dl_m(x - y) \,\fd{G}{y} \,.
\end{equation}
\end{defn}

One proves: Since $F$ and $G$ satisfy the wave front set property \eqref{eq:wavy-front} 
the pointwise product of distributions in 
\eqref{eq:Poisson-br} exists, and $\{F,G\}$ again
satisfies this wave front set property, hence
$\{F,G\} \in \sF$. Obviously,  the bracket \eqref{eq:Poisson-br} is skew-symmetric, $\{G,F\} = - \{F,G\}$
and satisfies the \emph{Leibniz rule}, $\{F, GH\}= \{F, G\}H+G \{F, H\}$;
and one verifies that it fulfills the \emph{Jacobi identity}; hence, it is indeed a Poisson bracket.

\section{Deformation quantization of the free theory}\label{sec:defo-quant}

Deformation quantization is mainly due to the paper \cite{BayenFFLS78}, which relies on much earlier 
work of John von~Neumann and on Gerstenhaber's algebraic deformation theory, 
and it gives an axiomatic formulation of the heuristic quantization formulas found by Weyl,
Groenewold and Moyal.
Wheras these works deal with quantum mechanics 
(i.e., finite dimensional systems), we apply it here to QFT.

\paragraph{Definition and properties of the star product.} Let $\sF_\hbar$ be the space of formal polynomials in $\hbar$
with coefficients in $\sF$.
The star product $\star\equiv\star_\hbar : \sF_\hbar \x \sF_\hbar \longto \sF_\hbar$ is a \emph{deformation
of the classical product} \eqref{eq:class-product} with deformation parameter $\hbar$, which is required to be
\begin{enumerate}
\item[(a)] \emph{bilinear} in its arguments;
\item[(b)] \emph{associative}; and, for $F,G\sim \hbar^0$; should satisfy:
\item[(c)]
$F \star_\hbar G \to F\.G$ (the classical product) as $\hbar \to 0$; and
\item[(d)]
$(F \star_\hbar G - G \star_\hbar F)/i\hbar \to \{F,G\}$ (the Poisson
bracket of the free theory) as $\hbar \to 0$.
\end{enumerate}

\begin{defn}
\label{df:star-product}
Given a suitable two-point function $H\in\sD'(\bR^d)$, the star product is defined by
\be\label{eq:general-star}
F \star_\hbar G :=\sum_{n=0}^\infty \frac{\hbar^n}{n!}
\int   dx_1 \cdots dx_n\,dy_1 \cdots dy_n
\,\,\frac{\dl^n F}{\dl\vf(x_1)\cdots\dl\vf(x_n)}
\prod_{l=1}^n H(x_l - y_l) \,
\frac{\dl^n G}{\dl\vf(y_1)\cdots\dl\vf(y_n)} \,.
\ee
Note that the sum over $n$ is \emph{finite} since $F,G\in\sF_\hbar$ are polynomials in $\vf$.
\end{defn}

The two-point function $H\in\sD'(\bR^d)$ should satisfy the following properties:
\begin{enumerate}
\item[(i)] The wave front set of $H$ should be such that the pointwise products of distributions 
appearing in \eqref{eq:general-star} exist and that $F\star G$ satisfies again the wave front set condition \eqref{eq:wavy-front} 
(for all $F,G\in\sF_\hbar$);
\item[(ii)] the above requirement (d) is satisfied iff the antisymmetric part of $H$ is given by
$\tfrac{1}{i}\bigl(H(z)-H(-z)\bigr)=\Dl_m(z)\ ;$ we also require
\item[(iii)] Lorentz invariance,
$H(\La z) = H(z)$ for all $\La \in \sL_+^\up$,
\item[(iv)] that $H$ is a solution of the free field equation, $(\square +m^2)H=0$,
\item[(v)] and that $\ovl{H(x)}=H(-x)$, which is equivalent to $(F\star G)^*=G^*\star F^*$.
\end{enumerate}

Due to (ii) and (iv), $H\equiv H_m$ depends on the mass $m\geq 0$ appearing in the free field equation;
hence, this holds also for the star product -- sometimes we signify this by writing $\star_m$ (instead of $\star$ or $\star_\hbar$).
The above requirements (a) and (c) are obviously satisfied and, with some effort, one can prove
associativity \cite{Waldmann07}.

The most obvious solution of the above requirements on $H$ is
the Wightman two-point function, $H_m=\Dl^+_m$. However, in even dimensions $d$, $\Dl^+_m$ is not smooth
in $m\geq 0$. The latter property can be reached using a Hadamard function instead, that is,
\be \label{eq:Hadamard}
H_m(x)=H_m^\mu(x)=\Dl^+_m(x)-m^{d-2}\,f_d(m^2x^2)\,\log(m^2/\mu^2)\ ,
\ee
where $\mu >0$ is a mass parameter and $f_d:\bR\to\bR$ is a certain analytic function (depending on the dimension $d$), hence
$\WF(H_m^\mu)=\WF(\Dl_m^+)$. That is,  $H_m^\mu$ solves the above requirement (i); obviously it also solves (ii), (iii)
and (v); and, since $(\square_x +m^2)f_d(m^2x^2)=0$, it also solves (iv). 
Note that $\Dl_m^+$ scales homogeneously, i.e., $\rho^{d-2} \Dl^+_{m/\rho}(\rho x) = \Dl^+_m(x)$,
but $H^\mu_m$ scales only \emph{almost homogeneously}, i.e., homogeneously up to logarithmic terms -- see Def.~\ref{df:scaling}.

\paragraph{States.} By definition, a \emph{state} $\om$ on the algebra $(\sF_\hbar,\star)$ is a functional 
$\om:(\sF_\hbar,\star)\longto\bC$ which is \emph{linear}, \emph{real} (i.e., $\om(F^*)=\ovl{\om(F)}$),
\emph{positive} (i.e., $\om(F^*\star F)\geq 0$) and \emph{normalized} (i.e., $\om(1)=1$).
Note that $\om$ itself may be a formal polynomial in $\hbar$; but, in $\om(F)$ (with $F\in\sF_\hbar$) the sum over the powers 
of $\hbar$ is an ordinary sum, in order that $\om(F)$ is a complex number (depending on $\hbar$).

A simple, but important, example is the \emph{vacuum state}:
\be\label{eq:vacuum}
\om_0(F) := f_0\word{where} F = f_0 + \sum_{n\geq 1}\langle f_n,\vf^{\ox n}\rangle\ .
\ee
For $H=\Dl_m^+$ one can prove that $\om_0$ is indeed positive, by using that 
$\int dx\,dy\,\,\ovl{g(x)}\,\Dl_m^+(x-y)\,g(y)\geq 0$ for all $g\in\sD(M,\bC)$; but the latter property may be 
violated for $H=H_m^\mu$.

\paragraph{On-shell fields.}
Introducing the space of solutions of the free field equation 
$$
\sC_0\equiv \sC_0^{(m)}:=\set{h\in\sC\,\big\vert\,(\square+m^2)h(x)=0}\ ,
$$
we define the space of \emph{on-shell fields} to be
$$
\sF_0^{(m)}:=\set{F_0:=F\bigr|_{\sC_0^{(m)}}\,\big\vert\,F\in\sF}\ .
$$
This definition is motivated by the fact that $\vf_0(x):=\vf(x)\bigr|_{\sC_0}$ satisfies the free field equation.

By using that $(\square +m^2)H=0$, one verifies that the star product on $\sF_\hbar$ induces a well-defined product on 
$$
\sF_{0,\hbar}^{(m)}:=\sF_\hbar\big\vert_{\sC_{S_0}^{(m)}}\word{by setting}F_0 \star G_0 := (F \star G)_0\ .
$$

Quantizing with $\Dl_m^+$, on-shell fieds $F_0\in\sF_{0,\hbar}^{(m)}$ may be identified with linear operators on Fock space:
\begin{thm}\label{th:star-Fock}
Let $\vf^\op(x)$ be the free, real scalar field $($for a given mass~$m)$
on the Fock space~$\gF$. % and let $\sF_{0,\hbar}^{(m)}:=\sF_\hbar\big\vert_{\sC_{S_0}}$.
Then the map
\be\label{eq:star-Fock}
\Phi\: \sF_{0,\hbar}^{(m)} \longto \Phi(\sF_{0,\hbar}^{(m)})
\subset \{\text{linear operators on } \gF\}
\ee
given by
$$
F_0=\sum_{n=0}^N \langle f_n, \vf_0^{\ox n}\rangle
\longmapsto \Phi(F_0)=\sum_{n=0}^N  \int dx_1 \cdots dx_n\,
\wick:\vf^\op(x_1) \cdots \vf^\op(x_n): \,f_n(x_1,\dots,x_n)
$$
(where $\wick: - :$ denotes normal ordering of Fock space operators) is an algebra isomorphism
for the star product on the left and the operator product on the right (i.e., $\Phi(F_0\star G_0)=\Phi(F_0)\,\Phi(G_0)$)
and also for the classical product on the left and the normally ordered product on the right
(i.e., $\Phi(F_0 \. G_0)=\wick:\Phi(F_0)\,\Phi(G_0):$). In addition, $\Phi$ 
respects the $*$-operation:
\be\label{eq:*op-Fock}
\big\langle\psi_1,\Phi(F_0^*)\psi_2\big\rangle_{\gF}=
\big\langle\Phi(F_0)\psi_1,\psi_2\big\rangle_{\gF}
\qquad\forall F_0\in\sF_{0,\hbar}^{(m)}
\ee
and for all $\psi_1,\psi_2$ in the domain of $\Phi(F_0)$ or $\Phi(F_0^*)$, respectively. 
\end{thm}

\section{Perturbative QFT}

Let $L_\intr=\sum_{k=1}^\infty L_k\,\ka^k\in\sP\pw{\ka}$ be the interaction Lagrangian, where $\ka$ is the coupling constant. 
With that,
\be\label{eq:S(g)}
S\equiv S(g):=\int dx\,\,\sum_{k=1}^\infty \bigl(g(x)\,\ka\bigr)^k L_k(x)\in\sF_\loc\pw{\ka}\ ,\quad g\in\sD(M)\ ,
\ee
is the \emph{adiabatically switched off interaction}. The main aim is to construct the pertinent scattering matrix ($S$-matrix), 
for which we make the ansatz
\be\label{eq:S-matrix}
\nS(S)=1+\sum_{n=1}^\infty \frac{i^n}{n!\hbar^n}\,T_n(S^{\ox n})\in\sF\pw{\ka}\ ,
\ee
which is a formal Laurent series in $\hbar$,
where $T_n:(\sF_\loc)^{\ox n}\to\sF$ is the time-ordered product (to $n$-th order), defined axiomatically in the following section.

\subsection{Axioms for the $T$-product}\label{sec:T-axioms}

In view of the inductive construction of $T=(T_n)$ we split the axioms into `basic axioms' and `renormalization conditions'.
The first basic axiom is
\begin{enumerate}
\item[(1)] \textbf{Linearity:}\index{basic axioms!linearity}
We require that
\be\label{eq:T-linear}
T_n \: \sF_\loc^{\ox n} \longrightarrow \sF \word{be linear.}
\ee
Note that here both the arguments and the values of $T_n$ are
\emph{off-shell fields}. 
\end{enumerate}

Our construction of the $T$-product is an inductive construction of the $\sF$-valued distributions
$T_n\bigl( A_1(x_1),\ldots, A_n(x_n)\bigr)\in\sD'(M^n,\sF)$ (for all
$A_1,\ldots,A_n\in\sP$), which should be connected to the maps 
$T_n:\sF_\loc^{\ox n} \to \sF$ \eqref{eq:T-linear} by
\be\label{eq:T(A(x))}
 \int dx_1\cdots dx_n\,\,
T_n\bigl( A_1(x_1),\ldots , A_n(x_n)\bigr)\,g_1(x_1) \cdots g_n(x_n)=
T_n\bigl( A_1(g_1) \oxyox A_n(g_n) \bigr)
\ee
for all $g_1,\dots,g_n \in \sD(\bM) $. But there is a problem with this fomula \eqref{eq:T(A(x))}: Since
$\int dx\,\,\del_x\bigl(g(x)\,A(x)\bigr)=0$, the relation \eqref{eq:T(A(x))} and Linearity of $T_n$ imply that
$$
\int dx\, \Bigl((\del g)(x)\, T_n\bigl(\ldots ,A(x), \ldots\bigr)
+g(x)\, T_n\bigl(\ldots ,(\del A)(x), \ldots\bigr)\Bigr)=0\ ;
$$
hence the \emph{Action Ward Identity} (AWI) must hold, that is,
\be\label{eq:AWI}
\text{\textbf{AWI:}}\quad\del_{x_l} T_n\bigl( \ldots , A(x_l),\ldots\bigr)
= T_n\bigl(\ldots ,\del_{x_l} A(x_l), \ldots\bigr),\quad \forall
A\in\sP\ ,\,\,1\leq l\leq n\ .
\ee
To define $T_n\bigl( A_1(x_1),\ldots, A_n(x_n)\bigr)$ in terms of the map 
$T_n:\sF_\loc^{\ox n} \to \sF$ in accordance with the AWI, we use Prop.~\ref{prop:balanced}: For balanced fields
$B_1,\dots,B_n\in\sP_\bal$ we define $T_n\bigl( B_1(x_1),\ldots , B_n(x_n)\bigr)$ by the formula \eqref{eq:T(A(x))},
and then, for arbitrary $A_1,\dots,A_n \in \sP$, we define
$T_n\bigl( A_1(x_1),\ldots , A_n(x_n) \bigr)$ by first
writing $A_i = \sum_{a_i} \del^{a_i}B_{ia_i}$ where $B_{ia_i} \in \sP_\bal$ and setting
\be\label{eq:def-AWI}
T_n\bigl(A_1(x_1),\ldots , A_n(x_n) \bigr):=\sum_{a_1,\dots,a_n}\del_{x_1}^{a_1} \cdots \del_{x_n}^{a_n}\,
T_n\bigl( B_{1a_1}(x_1),\ldots , B_{na_n}(x_n)\bigr)\ .
\ee
One easily verifies that with this definition the AWI \eqref{eq:AWI} holds for arbitrary $A\in\sP$ and that also the relation 
 \eqref{eq:T(A(x))} holds for all $A_1,\ldots,A_n\in\sP$.
 
The further basic axioms are:
 \begin{enumerate}
\item[(2)]  \textbf{Initial condition:}
$T_1(F) = F$ for any $F \in \sF_\loc\,$;
\item[(3)] \textbf{Symmetry:}
$T_n(F_{\pi(1)} \oxyox F_{\pi(n)}) = T_n(F_1 \oxyox F_n)
\quad\forall \,F_1,\ldots,F_n\in\sF_\loc \,$\\
and for all permutations $\pi$;
\item[(4)]  \textbf{Causality:}
For all $A_1,\ldots,A_n\in\sP$, $T_n$ fulfills the causal factorization:
\be\label{eq:Tprods-causal}
T_n\bigl(A_1(x_1),\dots,A_n(x_n)\bigr)
= T_k\bigl(A_1(x_1),\dots,A_k(x_k)\bigr)\star_m T_{n-k}\bigl(A_{k+1}(x_{k+1}),\dots, A_{n}(x_{n})\bigr)
\nonumber
\ee
whenever
$\{x_1,\dots,x_k\} \cap
\bigl( \{x_{k+1},\dots,x_n\} + \ovl V_- \bigr)
= \emptyset\,$;
\end{enumerate}
Assuming validity of axiom (3), one proves that axiom (4) is equivalent to the following causality relation for the $S$-matrix:
\be\label{eq:causality-S}
\nS(H + G + F) = \nS(H + G) \star \nS(G)^{\star-1} \star \nS(G + F)
\word{if} \supp H \cap (\supp F + \ovl V_-) = \emptyset\ ,
\ee
where $(-)^{\star-1}$ means the inverse w.r.t.~the star product.

Due to the axioms (1) (Linearity) and (3) (Symmetry), the formula \eqref{eq:S-matrix} implies that
$$
\nS^{(n)}:=\nS^{(n)}(0)=\tfrac{i^n}{\hbar^n}\,T_n \word{($n$th derivative of $\nS(F)$ at $F=0$), i.e., 
$T_n(F^{\ox n})= \bigl(\frac{\hbar}{i}\bigr)^n\frac{d^n}{d\la^n}\bigr|_{\la=0} \nS(\la F)$.}
$$

\medskip

\noindent We turn to the renormalization conditions:
\begin{enumerate}
\item[(5)] \textbf{Field independence:} $\dl T_n/\dl\vf=0$, that is, 
$\fd{T_n(F^{\ox n})}{x}=n\,T_n\Bigl(F^{\ox(n-1)}\ox\fd{F}{x}\Bigr)$ for all $F\in\sF_\loc$.
\end{enumerate}

\noindent Performing a (finite) Taylor expansion of $T_n\bigl(A_1(x_1),\dots,A_n(x_n)\bigr)$ in $\vf$ with respect to $\vf=0$, one shows 
that Field Independence is equivalent to the validity of the \emph{causal Wick expansion}:  For \emph{monomials} 
$A_1,\dots,A_n\in\sP$ it holds that
\be\label{eq:causal-Wick-expan}
T_n\bigl( A_1(x_1),\dots, A_n(x_n) \bigr)
= \sum_{\unl A_j \subseteq A_j}\! \om_0\Bigl(
T_n\bigl( \unl A_1(x_1),\dots, \unl A_n(x_n) \bigr) \Bigr)\,
\ovl A_1(x_1) \cdots \ovl A_n(x_n),
\ee
where each \emph{submonomial} $\unl A$ of a given
monomial $A$ (i.e., $\unl A\subseteq A$) and its \emph{complementary submonomial} $\ovl A$ are defined by
\be
\unl A:= \frac{\del^k A}{\del(\del^{a_1}\vf)\cdots\del(\del^{a_k}\vf)}
\neq 0,\qquad
\ovl A:= C_{a_1\dots a_k} \,\del^{a_1}\vf \cdots \del^{a_k}\vf
\label{eq:submonomials} 
\ee
(no sum over $a_1,\dots,a_k$ in the formula for $\ovl A$), with $C_{a_1\dots a_k}$ being a certain combinatorial factor.
The range of the sum $\sum_{\unl A \subseteq A}$ are all
$k\in\bN$ and $a_1,\dots,a_k\in\bN^d$ which yield a $\unl A\not= 0$.
(For $k = 0$ we have $\unl A = A$ and $\ovl A = 1$.) The main message of the causal Wick expansion is that
$T_n\bigl( A_1(x_1),\dots \bigr)\in\sD'(M^n,\sF)$ is uniquely determined by the family of $\bC$-valued distributions
$\om_0\bigl(T_n\bigl( \unl A_1(x_1),\dots \bigr) \bigr),\,\,\unl A_j \subseteq A_j$.

\begin{enumerate}
\item[(6)] \textbf{Unitarity and field parity:}
In order that $\nS(S)$ is unitary for real interactions (i.e., $S=S^*$), we require
\be\label{eq:unitarity}
\nS(F)^*=\nS(F^*)^{\star-1}\quad\forall F\in\sF_\loc .
\ee
Field parity is the condition
\be\label{eq:field-parity}
\al\circ T_n=T_n\circ\al^{\ox n},\word{where $\al:\sF\to\sF$ is defined by} (\al F)[h]:=F[-h]\quad\forall h\in\sF.
\ee

\item[(7)] \textbf{Poincar\'e covariance:} $\bt_{\La,a}\circ T_n= T_n\circ\bt_{\La,a}^{\,\,\ox n}\quad\forall
(\La,a)\in \sP_+^\up$,\\
where $\bt_{\La,a}:\sF\to\sF$ is defined by 
$$
\bt_{\La,a}\sum_n\langle f_n,\vf^{\ox n}\rangle:=\sum_n\langle f_n(x_1,\ldots,x_n),\vf(\La x_1 + a)\oxyox\vf(\La x_n + a)\rangle.
$$
\end{enumerate}

\noindent Considering only Translation covariance (i.e., $\La = 1$), we
conclude that the $\bC$-valued distributions
\begin{equation}
t_n(A_1,\dots,A_n)(x_1 - x_n,\dots, x_{n-1} - x_n)
:= \om_0\Bigl( T_n\bigl( A_1(x_1),\dots, A_n(x_n) \bigr) \Bigr)
\label{eq:t-def} 
\end{equation}
depend \emph{only} on the relative coordinates, since
$\om_0 \circ \bt_{\La,a} = \om_0$. By
using the causal Wick expansion \eqref{eq:causal-Wick-expan}, one
easily verifies that this property \eqref{eq:t-def} is in fact \emph{equivalent} to Translation covariance.

\begin{enumerate}
\item[(8)] \textbf{Off-shell field equation:}
\begin{align}\label{eq:FE-T}
T_n\bigl(\vf(g) \ox &  F_1\oxyox F_{n-1}\bigr)=\vf(g)\,\,
T_{n-1}\bigl(F_1\oxyox F_{n-1}\bigr)\\
&+\hbar\int dx\,dy\,\,g(x)\,H^F_m(x-y)\,\sum_{k=1}^{n-1}
T_{n-1}\bigl(F_1\oxyox \fd{F_k}{y}\oxyox F_{n-1}\bigr)\ ,\nonumber
\end{align}
where $g\in\sD(\bM)$ and $H^F_m$ is the Feynman propagator belonging to the two-point function $H_m$, that is,
$H_m^F(x):=\th(x^0)\,H_m(x)+\th(-x^0)\,H(-x)$.

\item[(9)] \textbf{Smoothness in the mass $m\geq 0$:}
By the basic axioms, $T\equiv T^{(m)}$ depends on the mass $m$ of the free field equation via the star product
$\star_m$ appearing in the Causality axiom. We require that the distributions
$$
t_n^{(m)}(A_1,\dots,A_n) \word{\eqref{eq:t-def} depend smoothly on $m\geq 0$, for all $A_1,\ldots,A_n\in\sP$ and all $n$.}
$$
This axiom excludes quantization with $\Dl^+_m$; hence, in the following we quantize with a Hadamard function $H_m^\mu$.
\end{enumerate}

\noindent The next axiom deals with the scaling behaviour of the $T$-product. For this and also in view of the 
inductive construction of the $T$-product we introduce some notions:
 \begin{defn}\label{df:scaling} 
 Let $f\in\sD'(\bR^k)$ or $f\in\sD'(\bR^k \less \{0\})$.
 \begin{enumerate}
 \item[(a)] We say that $f$ scales \emph{almost homogeneously} with
\emph{degree} $D \in \bC$ and \emph{power} $N \in \bN$ iff
 \be\label{eq:almost-homog} 
\Bigl(\bE_k + D\Bigr)^{N+1} f(y_1,\dots,y_k) = 0 \word{and}
\Bigl(\bE_k + D\Bigr)^N f(y_1,\dots,y_k) \neq 0\ ,
\ee
where $\bE_k := \sum_{r=1}^k y_r\,\del/\del y_r$ is the Euler operator.
When $N = 0$, we say there is \emph{homogeneous} scaling of degree~$D$.
\item[(b)] For $f(y)=f^{(m)}(y)$ (where $y\in\bR^k$) being differentiable in the mass $m\geq 0$, we say that 
$f^{(m)}(y)$ scales \emph{almost homogeneously under $(y,m)\mapsto (\rho y,m/\rho)$}
with \emph{degree} $D \in \bC$ and \emph{power} $N \in \bN$ iff the relations \eqref{eq:almost-homog} hold
for $(\bE-m\,\del/\del m)$ in place of $\bE$.
\item[(c)] The \emph{scaling degree} (with respect to the origin) of $f$ is given by
$$
\sd(f) := \inf\set{r \in \bR\,
\,\big\vert\, \lim_{\rho\downto 0} \rho^r\,f(\rho y) = 0}\ ,
$$
where $\inf\,\emptyset:=\infty$ and $\inf\,\bR:=-\infty$.
\end{enumerate}
\end{defn}
\noindent For example,%
\footnote{$\dl_{(k)}$ denotes the $\dl$-distribution supported at the origin of $\bR^k$.}
 $\del^a\dl_{(k)}\in\sD'(\bR^k)$ scales homogeneously with degree $D=k+|a|$. 
Obviously, a distribution $f$ scaling almost homogeneously with
degree $D \in \bC$ (and arbitrary power $N$) has scaling degree $\sd(f)=\mathrm{Re}\, D$.

We also need the ``mass dimension'' of a \emph{monomial} $A\in\sP$.
\begin{defn}\label{df:mass-dim}
The \emph{mass dimension} of $\del^a\vf \in \sP$ is defined by
$$
\dim \del^a\vf := \frac{d-2}{2} + |a| \word{for} a\in\bN^d\ .
$$
For \emph{monomials} $A_1, A_2 \in \sP$, we agree that
$\dim (A_1A_2) := \dim A_1 + \dim A_2$.
\end{defn}
Denoting by $\sP_j$ the vector space spanned by all monomials $A \in \sP$
with $\dim A = j$ we introduce the \emph{set} of ``homogeneous'' polynomials:
$\sP_\homog := \bigcup_{j} \sP_j$ (where $j\in\bN/2$ if $d$ is odd, and $j\in\bN$ otherwise).

\begin{enumerate}
\item[(10)] \textbf{Scaling:}  For all \emph{monomials} $A_1,\dots,A_{n}\in\sP$ and all $n\geq 2$ we require that
\begin{align*}
t_n^{(m)}(A_1,\ldots,A_n)(y)&\word{scales almost homogeneously under $(y,m)\mapsto (\rho y,m/\rho)$ with degree}\\ 
&\word{$D=\sum_{l=1}^n\dim A_l$ and an arbitrary power $N<\infty$ (where $y\in\bR^{d(n-1)}$).} 
\end{align*}

\item[(11)] \textbf{\boldmath $\hbar$-dependence:} For all \emph{monomials} $A_1,\dots,A_{n}\in\sP$ 
fulfilling $A_j\sim \hbar^0\,\,\forall j$ and all $n\geq 2$, we require
$$
t_n(A_1,\ldots,A_n)\sim\hbar^{\sum_{j=1}^n |A_j|/2}\ ,
$$
where $|A|$ is the degree of the monomial $A$, i.e., $A(x)[\la h]=\la^{|A|}\, A(x)[h]$ for all $h\in\sC,\,\la>0$.
\end{enumerate}

\noindent We point out that in axiom (10) the degree $D$ fulfills $D\in\bN$ (also in odd dimensions $d$), and in axiom (11) 
the power of $\hbar$ satisfies $\sum_{j=1}^n |A_j|/2\in\bN$. Both statements rely on the observation that Field parity 
\eqref{eq:field-parity} and $\om_0\circ\al =\om_0$ imply that 
$$
t_n(A_1,\ldots,A_n)=0\word{if}\sum_{j=1}^n |A_j| \word{is odd;}
$$
and, for the statement about the degree, we also use that for $A=c\prod_{j=1}^J\del^{a_j}\vf$ ($c\in\bR$) it holds that
$\dim A=J\cdot\dim\vf+\sum_{j=1}^J|a_j|=|A|\cdot\tfrac{d-2}2+\sum_{j=1}^J|a_j|$.

\subsection{Inductive construction of the $T$-product}
The $T$-product $T=(T_n)$ is constructed by induction on $n$, starting with axiom (2) (Initial condition). Turning to the inductive step
$(n-1)\to n$ we introduce the thin diagonal in $M^n$:
\be\label{eq:Dln}
\Dl_n:=\set{(x_1,\ldots,x_n)\in M^n\,\big\vert\,x_1=\ldots =x_n}\ .
\ee

\paragraph{Inductive step, off the thin diagonal $\Dl_n$.}
The axioms (4) (Causality) and (3) (Symmetry) imply the following: 
For each point $x\in M^n\less\Dl_n$ there exist a neighbouhood $U_x\subset M^n$ of $x$, a $k\in\{1,\ldots, n-1\}$ and a 
permutation $\pi$ such that $\{x_{\pi(1)},\dots,x_{\pi(k)}\} \cap
\bigl( \{x_{\pi(k+1)},\dots,x_{\pi(n)}\} + \ovl V_- \bigr)= \emptyset\,$ for all $(x_1,\ldots,x_n)\in U_x$, hence
\begin{align*}
T_n\bigl(A_1(x_1),\dots, A_n(x_n) \bigr)=\,&\,T_k\bigl( A_{\pi(1)}(x_{\pi(1)}),\dots, A_{\pi(k)}(x_{\pi(k)}) \bigr)\\
&\star T_{n-k}\bigl( A_{\pi(k+1)}(x_{\pi(k+1)}),\dots, A_{\pi(n)}(x_{\pi(n)}) \bigr)\word{on $\sD(U_x)$}
\end{align*}
for all $A_1,\ldots,A_n\in\sP$; that is,
$T_n\bigl( A_1(x_1),\dots\bigr)\vert_{\sD(U_x)}$ is uniquely determined in terms of the inductively known
$(T_k)_{1\leq k <n}$. This observation gives part (a) of the following Theorem.

\begin{thm}\label{th:T^0} 
Let $(T_k)_{1\leq k <n}$ be constructed.
\begin{enumerate}
\item[(a)] Uniqueness: If there exists some map $T_n:\sP^{\ox n}\to\sD'(M^n,\sF)$ fulfilling the basic axioms, then its image is
uniquely determined on $\sD(M^n\less\Dl_n)$.
\item[(b)] Existence: There exists a map $T_n^0:\sP^{\ox n}\to\sD'(M^n\less\Dl_n,\sF)$ satisfying all axioms (1)-(11).
\end{enumerate}
\end{thm}

The proof of part (b) is constructive, using a partition of unity subordinate to an open cover of $M^n\less\Dl_n$ \cite{Stora08,BF00}.

Alternatively, the use of a partition of unity in the inductive construction of $T_n^0$ can be avoided by working with
the distribution splitting method of Epstein and Glaser \cite{EpsteinG73, Scharf}, on the price of a more
complicated combinatorics.%
\footnote{Note that for the construction of $T_n^0$ the distribution splitting can be done by multiplication with a 
Heaviside-function. Hence, the distribution splitting problem in the Epstein--Glaser construction is an equivalent reformulation of
the extension problem $T_n^0\to T_n$ treated in the next paragraph.}

\paragraph{Extension to the thin diagonal $\Dl_n$.} The extension of $T_n^0$ (taking values in $\sD'(M^n\less\Dl_n,\sF)$)
to a well-defined $T_n$ (taking values in $\sD'(M^n,\sF)$) is non-unique and it corresponds
to what is called ``renormalization'' in conventional approaches.
By part (a) of Thm.~\ref{th:T^0}, the renormalization conditions are
not used for the construction of $T_n^0$, but they give guidance on how to do this extension
and reduce the non-uniqueness drastically.

Since we are constructing $T_n:\sP^{\ox n}\to\sD'(M^n,\sF)$ (instead of $T_n:\sF_\loc^{\ox n}\to\sF$), the AWI plays 
the role of an additional renormalization condition. We fulfil it by first constructing
$T_n\bigl(B_1(x_1),\ldots,B_n(x_n)\bigr)$ for all $B_1,\ldots ,B_n\in\sP_\bal$ and then we define
 $T_n\bigl(A_1(x_1),\ldots,A_n(x_n)\bigr)$ for arbitrary $A_1,\ldots ,A_n\in\sP$ by \eqref{eq:def-AWI}.
 
 By Linearity (axiom (1)), the causal Wick expansion (axiom (5)) 
 and Translation covariance (axiom (7)), the problem is simplified to the extension of
 $$
 t^0(y)\equiv t^{(m),0}_n(B_1,\ldots,B_n)(y):=\om_0\Bigl( T_n^{(m),0}\bigl( B_1(x_1),\dots, B_n(x_n) \bigr) \Bigr)
 \in\sD'(\bR^{d(n-1)}\less\{0\},\bC),
 $$
 where $y:=(x_1-x_n,\ldots,x_{n-1}-x_n)$ and $B_1,\ldots ,B_n\in\sP_\bal\cap\sP_\homog$, to a distribution 
 $t\in\sD'(\bR^{d(n-1)},\bC)$. In view of axiom (10) (Scaling) we require here and in the following that $B_j\in\sP_\bal\cap\sP_\homog$; 
 this is sufficient since there exists a basis of $\sP_\bal$ lying in $\sP_\homog$.

Obviously, any extension $t\in\sD'(\bR^k)$ of a given $t^0\in\sD'(\bR^k\less\{0\})$ obeys $\sd(t) \geq \sd(t^0)$. 
Looking for extensions which do not increase the scaling degree, existence and uniqueness of this problem is 
answered by the following \cite{BF00}:
\begin{thm}\label{th:extens-exist}
Let $t^0 \in \sD'(\bR^k \less \{0\})$. Then:
\begin{enumerate}
\item[\textup{(a)}]
If $\sd(t^0) < k$, there is a \emph{unique} extension (called ``direct extension'')
$t \in \sD'(\bR^k)$ fulfilling the condition $\sd(t) = \sd(t^0)$.
\item[\textup{(b)}]
If $k \leq \sd(t^0) < \infty$, there are \emph{several} extensions
$t \in \sD'(\bR^k)$ satisfying the condition $\sd(t) = \sd(t^0)$. In
this case, given a particular solution $t_0$, the general solution is
of the form
\be\label{eq:general-extension}
t = t_0 + \sum_{|a|\leq \sd(t^0)-k} C_a \,\del^a\dl_{(k)}
\word{with}  C_a \in \bC.
\ee
\end{enumerate}
\end{thm}
The proof is constructive, the idea of the construction is given in Sect.~\ref{sec:techniques}.

Turning to the maintenance of the axioms (9) (Smoothness in $m\geq 0$) and (10) (Scaling) in the extension 
$ t^0\equiv t^{(m),0}_n(B_1,\ldots,B_n)\to t$ (where $B_1,\ldots ,B_n\in\sP_\bal\cap\sP_\homog$), we set
$D:=\sum_{j=1}^n \dim B_j\in\bN$ and
we first point out that in the case $m=0$ we may apply the following \cite{Hormander90,HollandsW01-05,NikolovST14}:
\begin{prop}
\label{pr:extens-exist}
Let $t^0 \in \sD'(\bR^k \less \{0\})$ scale almost homogeneously with
degree $D \in \bC$ and power $N_0 \in \bN$. Then there exists an
extension $t \in \sD'(\bR^k)$ which scales also almost homogeneously
with degree~$D$ and power $N \geq N_0$:
\begin{enumerate}
\item[\rm (i)]
If $D \notin \bN + k$, then $t$ is unique and $N = N_0$;
\item[\rm (ii)]
if $D \in \bN + k$, then $t$ is non-unique and $N = N_0$ or
$N = N_0 + 1$. In this case, given a particular solution $t_0$, the
general solution is of the form
\be\label{eq:general-homogeneous-extension}
t = t_0 + \sum_{|a|=D-k} C_a \,\del^a\dl_{(k)}
\word{with arbitrary}  C_a \in \bC.
\ee
\end{enumerate}
For the subcase $\mathrm{Re}\,D\,(=\sd(t^0)) < k$ of case~{\rm(i)},
the unique $t$ agrees with the direct
extension of~$t^0$ of part (a) of Thm.~\ref{th:extens-exist}.
\end{prop}

For $m>0$, Smoothness in $m\geq  0$ of $t^{(m),0} \in\sD'(\bR^{d(n-1)}\less\{0\},\bC)$ ensures the existence of the 
Taylor expansion
\begin{align}\label{eq:massy-Taylor}
t^{(m),0}(y) = \sum_{l=0}^{D-d(n-1)} \frac{m^l}{l!}\, u_l^0(y)
+ m^{D-d(n-1)+1} t_\red^{(m),0}(y)
\word{with}
u_l^0(y) := \frac{\del^l t^{(m),0}(y)}{\del m^l} \biggr|_{m=0}.
\end{align}
Almost homogeneous scaling under $(y,m)\mapsto (\rho y,m/\rho)$ of $t^{(m),0}$ with degree $D$ implies 
almost homogeneous scaling under $y\mapsto \rho y$ of $u_l^0$ with degree $D-l\in\bN+d(n-1)$, hence we may apply part (ii) of
Prop.~\ref{pr:extens-exist} for the extension $u_l^0\to u_l\in\sD'(\bR^{d(n-1)},\bC)$.
Considering $t_\red^{(m),0}$, one shows that the validity of the axioms (9) and (10) for $t^{(m),0}$ implies 
the validity of (9) for  $t_\red^{(m),0}$ and that $\sd(t_\red^{(m),0})<d(n-1)$. Hence, Thm.~\ref{th:extens-exist}(a) 
(i.e., the direct extension) provides a unique extension $t_\red^{(m)}$ with $\sd(t_\red^{(m)})= \sd(t_\red^{(m),0})$ . 
The latter maintains smoothness in $m\geq 0$ and one verifies that it also maintains 
almost homogeneous scaling under $(y,m)\mapsto (\rho y,m/\rho)$ with degree $d(n-1)-1$.

The construction given so far guarantees that the resulting $T_n$ satisfies the renormalization conditions AWI, 
Field independence, Translation covariance, Smoothness in $m\geq 0$ and Scaling.
How to maintain the further renormalization conditions in the extension $t_n^0\to t_n$?
\begin{itemize}
\item For axiom (10) ($\hbar$-dependence) this is reached by doing the extension 
in each order of~$\hbar$ individually.

\item Turning to the symmetries required in axiom (6) (Unitarity and Field parity) and (7) (Lorentz covariance) we 
quote the following general existence result: 
\emph{Given a $t_n^0$ scaling almost homogeneously and being symmetric w.r.t.~a certain group 
$\sG$, there exists an extension $t_n$ which scales also almost homogeneously and is also $\sG$-symmetric, 
if all finite-dimensional representations of $\sG$ are completely reducible.}
This assumption is satisfied for Unitarity and Field parity (in both cases $\sG$ is the group $(\{-1,1\},\,\cdot\,)$ and also 
for the Lorentz group $ \sL_+^\up$.

The \emph{construction} of a $\sG$-symmetric $t_n$ is an easy task for Unitarity and Field parity (one starts with
an extension satisfying all other renormalization conditions and symmetrizes it w.r.t.~$\sG$), but for
Lorentz covariance this requires some effort. (See also~\cite{PopineauS16,Scharf}).

\item The Off-shell field equation (axiom (8)) is satisfied by defining $t_n(\vf,B_1,\ldots,B_{n-1})$ 
(with $B_1,\ldots,B_{n-1}\in\sP_\bal\cap\sP_\homog$) in terms of $t_{n-1}$ by the 
vacuum expectation value of the relation \eqref{eq:FE-T} (note, that the first term on the r.h.s.~of \eqref{eq:FE-T}
does not contribute). The so defined $t_n(\vf,B_1,\ldots,B_{n-1})$ is an extension of
$t_n^0(\vf,B_1,\ldots,B_{n-1})$ (because the latter fulfills \eqref{eq:FE-T}) 
and one verifies easily that it satisfies all other renormalization conditions.
\end{itemize}

\subsection{Techniques to renormalize}\label{sec:techniques}

In this section we sketch main techniques to compute the extension $\sD'(\bR^k \less \{0\})\ni t^0\to t\in\sD'(\bR^k)$.

\paragraph{Direct extension and $W$-extension.} In the proof of Thm.~\ref{th:extens-exist} the extension $t$ is constructed as 
follows \cite{BF00,EpsteinG73}:
\begin{itemize}
\item For $\sd(t^0)<k$ the (unique) ``direct extension'' is obtained by 
\be\label{eq:direct-extension}
\duo< t, h> := \lim_{\rho\to\infty} \duo< t^0, \chi_\rho h>
\qquad\forall h\in\sD(\bR^k)\ ,
\ee
where  $\chi \in C^\infty(\bR^k)$ is such that $0 \leq \chi(x) \leq 1$,
$\chi(x) = 0$ for $|x| \leq 1$ and $\chi(x) = 1$ for $|x| \geq 2$ and we use the notation 
$\chi_\rho(x):=\chi(\rho x)$. Since $\chi_\rho(x)h(x)\in\sD(\bR^k \less \{0\})$ for any $\rho > 0$, 
the expression $\duo< t^0, \chi_\rho h>$ exists. One proves that the limit \eqref{eq:direct-extension}
exists and defines a distribution $t \in \sD'(\bR^k)$. Since for any $h_1\in\sD(\bR^k\less\{0\})$
it holds that $\chi_\rho h_1=h_1$ for $\rho$ sufficiently large, this $t$ is indeed an extension of $t^0$.
In addition one shows that $\sd(t) = \sd(t^0)$.

For practical computations the formula \eqref{eq:direct-extension} means that the direct extension $t$ is given by 
the same formula as $t^0$.

\item For $\sd(t^0)\geq k$: Let $\om := \sd(t^0) - k$ be the \emph{singular order}
of~$t^0$. Introducing the subspace of test functions
\begin{equation}
\sD_\om:=\sD_\om(\bR^k) := \set{h \in \sD(\bR^k)\,\big\vert\,
\del^a h(0) = 0 \text{ for } |a| \leq \om}\ ,
\label{eq:Domega} 
\end{equation}
one proves that \emph{$t^0$ has a unique extension $t_\om$ to $\sD'_\om$ satisfying
$\sd(t_\om) = \sd(t^0)$} -- roughly speaking, the direct extension applies also in this case:
\be
\duo< t_\om, h> = \lim_{\rho\to\infty} \duo< t^0, \chi_\rho h>\ .
\ee

Each projector $W\: \sD(\bR^k) \longto \sD_\om$ defines an extension $t^W \in \sD'(\bR^k)$
of~$t^0$ (called ``$W$-extension'') by
\begin{equation}
\duo< t^W, h> := \duo< t_\om, Wh>\,,
\label{eq:W-extension} 
\end{equation}
Since $Wh = h$ for $h \in \sD(\bR^k \less \{0\})$, the relations
$$
\duo< t^W, h> = \duo< t_\om, Wh> = \duo< t_\om, h> = \duo< t^0, h>
$$
show that $t^W$ is indeed an extension of~$t^0$. More elaborate is the proof of $\sd(t^W)=\sd(t^0)$.

Any set of functions $w_a \in \sD(\bR^k)$ (where
$a \in \bN^k$ with $|a| \leq \om$) satisfying
\be\label{eq:wa}
\del^b w_a(0) = \dl_a^b\quad\forall b\in\bN^k\ ,\,\,|b| \leq \om
\ee
defines such a projector~$W$ by
\begin{equation}
Wh(x) := h(x) - \sum_{|a|\leq \om} \del^a h(0)\,w_a(x).
\label{eq:W-projector} 
\end{equation}
One can prove that  every extension~$t$ having $\sd(t) = \sd(t^0)$ 
is a $W$-extension \eqref{eq:W-extension} with the projector $W$ given in
terms of a family of functions $(w_a)$ \eqref{eq:wa}
by~\eqref{eq:W-projector}.
\end{itemize}
 
For practical computations, the $W$-extension has essential disadvantages: Firstly, for $t^W$ (given
by \eqref{eq:W-extension} and \eqref{eq:W-projector})
Lorentz covariance is at least not manifest, since there does not exist any Lorentz covariant
$w_a \in \sD(\bR^k)$. Secondly, to compute $t^W$ explicitly, one needs explicit formulas for the functions $(w_a)$
which makes the computation unhandy.

Due to this and due to the Taylor expansion in $m\geq 0$ of $t^0$ \eqref{eq:massy-Taylor}, 
in praxis one is left with the following problem: \emph{Given a distribution
$t^0 \in \sD'(\bR^k \less \{0\})$ which scales almost homogeneously with
degree $D \in k+\bN$, find an extension $t \in \sD'(\bR^k)$ which
scales also almost homogeneously with degree~$D$.} We are going to sketch two techniques solving this problem.
 
\paragraph{Differential renormalization.} The idea is to trace back the case $D\geq k$ to the simple case
$D<k$, in which the solution is unique and
obtained by the direct extension (Prop.~\ref{pr:extens-exist}), in the following way: Write $t^0$ as a derivative
of a distribution $f^0\in \sD'(\bR^k \less \{0\})$ which scales almost
homogeneously with degree $D-l<k$, where $l\in\bN\less\{0\}$; more precisely
\begin{equation}
  t^0 =\mathfrak{D}f^0 \word{with}
\mathfrak{D}=\sum_{|a|=l}C_a\del^a \ ,\,\,\, C_a\in\bC\ ,
\word{and} (\bE_k +D-l)^N \,f^0=0\label{eq:t^0=Df^0}
\end{equation}
for $N\in\bN$ sufficiently large.
Let $f\in \sD'(\bR^k)$ be the direct extension of $f^0$; it scales also almost
homogeneously with the same degree~$D-l$ and the same power. Then
\begin{equation}
 t:= \mathfrak{D}f\label{eq:t=Df}
\end{equation}
exists in $\sD'(\bR^k)$ and
is an extension of $t^0$, because for $h\in\sD(\bR^k \less \{0\})$ we get
\be\label{eq:t-diffren-extens}
\duo<t,h>=\duo<\mathfrak{D}f,h>=(-1)^l\,\duo<f,\mathfrak{D}h>
=(-1)^l\,\duo<f^0,\mathfrak{D}h>=\duo<\mathfrak{D}f^0,h>=\duo<t^0,h>\ ,
\ee
and one easily verifies that $(\bE_k +D)^N \,t=0$.

In praxis, the difficult step is to find a distribution $f^0\in \sD'(\bR^k\less\{ 0 \})$
satisfying the conditions \eqref{eq:t^0=Df^0}. A general method to
solve this problem is not known; however, differential renormalization has been successfully applied to
a wealth of concrete examples, see e.g.~\cite{Elara,Prange99}.

\paragraph{Analytic renormalization.}%
\footnote{Analytic renormalization was first applied to $x$-space Epstein--Glaser renormalization by Hollands
\cite{Hollands08}. Essentially, we follow \cite{DuetschFKR14}.}
Roughly, the idea is to solve the above given extension problem 
as follows: One introduces a $\zeta$-dependent
``regularized'' distribution $t^{\zeta\, 0}\in\sD'(\bR^k\less\{0\})$
(where $\zeta\in\bC\less\{0\}$ and $|\zeta|$ sufficiently small),
such that $\lim_{\zeta\to 0}t^{\zeta\, 0}=t^0$
and that $t^{\zeta\, 0}$ scales almost homogeneously with a \emph{non-integer
degree} $D_\zeta$. From Prop.~\ref{pr:extens-exist} we know
that $t^{\zeta\, 0}$ has a unique extension $t^\zeta\in\sD'(\bR^k)$ which
scales almost homogeneously with the same degree $D_\zeta$. The explicit computation 
of $t^\zeta$ is generically much simpler than the
computation of a solution of the original extension task, mostly this can be done by means of 
differential renormalization -- this is the gain of the regularization. 
For the resulting extension $t^\zeta$ one then
removes the regularization, i.e., one performs the limit
$\zeta\to 0$. In order that this limit exists in
$\sD'(\bR^k)$, one has to subtract suitable local terms.

To explain this in detail, first note that almost homogeneous scaling of $t^0$
implies that the above introduced unique extension
$t_\om$ to $\sD'_\om(\bR^k)$ \eqref{eq:Domega} (where $\om:=D-k$) scales also almost homogeneously.

\begin{defn}[Analytic regularization]\label{def:regularization} 
With the given assumptions (see the above formulated problem)
and notations, a family of distributions $\{t^\zeta\}_{\zeta\in\Omega\less\{0\}}$,
$t^\zeta\in\sD'(\bR^k)$, with $\Omega\subseteq\bC$ a neighbourhood of the origin,
is called a \emph{regularization} of $t^0$, if
\be\label{eq:regularization}
\lim_{\zeta\to 0}\duo< t^\zeta,h>=\duo< t_\om ,h>
\quad\quad \forall h\in \sD_\om(\bR^k)\ ,
\ee
and if $t^\zeta$ scales almost homogeneously with degree
$D_\zeta=D+ D_1\,\zeta$ for some constant $D_1\in \bC\less\{0\}$. 
The regularization $\{t^\zeta\}$ is called \emph{analytic}, if for all $h\in\sD(\bR^k)$ the map
\be\label{eq:meromorph}
\Omega\setminus\{0\}\ni\zeta\longmapsto \duo< t^\zeta,h>
\ee
is analytic with a pole of finite order at the origin.
\end{defn}

Let $\{t^\zeta\}$ be an analytic regularization of $t^0$ and let $t$ be an almost homogeneous
extension of $t^0$. As mentioned above, $t$ can be written as a $W$-extension, that is, there exist functions $(w_a)$ 
\eqref{eq:wa} such that
\be\label{eq:W-zeta}
\duo<t,h>=\duo<t^W,h>=\duo<t_\om,Wh>=\lim_{\zeta\to 0}\duo< t^\zeta,Wh>=
\lim_{\zeta\to 0}\Bigl(\duo< t^\zeta,h>
- \sum_{|a|\leq \om} \duo<t^\zeta,w_a>\,\del^a h(0)\Bigr)\ ,
\ee
by using \eqref{eq:W-extension}, \eqref{eq:regularization} and finally \eqref{eq:W-projector}.
In general,  the limit of the individual terms on the right-hand side
might not exist. However, each
term can be expanded in a Laurent series around $\zeta=0$, and since the
overall limit is finite, the principal parts ($\pp$) of these Laurent series must cancel out:
\be\label{eq:pp=pp}
\big\langle\pp(t^\zeta),h\big\rangle:=\pp\bigl(\duo <t^\zeta,h>\bigr) = \sum_{|a|\leq \om}
\pp\bigl(\duo<t^\zeta,w_a>\bigr)\,\,\del^a h(0)
\ ,\quad \forall h\in\sD(\bR^k)\ .
\ee
We conclude that
\be\label{eq:pp(t-zeta)}
\pp\bigl(t^\zeta(x)\bigr) = \sum_{|a|\leq D-k}C_a(\zeta)\,\del^a\dl(x)\ ,\word{where}
C_a(\zeta)=(-1)^{|a|}\,\pp\bigl(\duo<t^\zeta,w_a>\bigr)\ .
\ee

\begin{prop}[Minimal subtraction]\label{pr:MS}
\begin{enumerate}
\item[\rm (a)] The sum in \eqref{eq:pp(t-zeta)} runs only over $|a|=D-k$, that is,
the principal part $\pp(t^\zeta)$ is a local distribution
which scales homogeneously with degree $D$.
\item[\rm (b)] The regular part $\,\rp(t^\zeta):=t^\zeta-\pp(t^\zeta)$ defines by
\be\label{eq:def-MS}
\duo< t^\MS,h> :=\lim_{\zeta\to 0}\, \rp\bigl(\duo< t^\zeta,h>\bigr)\ ,\quad
\forall h\in\sD(\bR^k)\ ,
\ee
a distinguished extension of $t^0$ which scales almost homogeneously with
degree $D$ (``minimal subtraction'').
\end{enumerate}
\end{prop}

That $t^\MS$ is an extension of $t^0$ with $\sd(t^\MS)=\sd(t^0)$ (i.e., Prop.~\ref{pr:MS} without the 
statements about almost homogeneous scaling), can easily be seen by proceeding as follows:
We compare $t^\MS$ with the initial extension $t=t^W$.
Using \eqref{eq:W-zeta}, \eqref{eq:pp=pp} and the definition of $t^\MS$ \eqref{eq:def-MS}, we obtain
\begin{align}\label{eq:t-tMS}
\duo<t,h>={}&\lim_{\zeta\to 0}\Biggl(\duo< t^\zeta,h>
- \sum_{|a|\leq \om} \Bigl(\pp\bigl(\duo<t^\zeta,w_a>\bigr)+
\rp\bigl(\duo<t^\zeta,w_a>\bigr)\Bigr)\,\,\del^a h(0)\Biggr)\nonumber\\
={}&\duo<t^\MS,h>-\sum_{|a|\leq \om}\duo<t^\MS,w_a>\,\,\del^a h(0)\ ,
\quad h\in\sD(\bR^k)\  .
\end{align}
Hence, $t^\MS$ differs from $t$ by a term of the form
$\,t^\MS-t=\sum_{|a|\leq \om}b_a\,\del^a\dl\ ,\,\,b_a\in\bC\ $.
More involved is the proof of the statements that $\pp(t^\zeta)$ and $t^\MS$ scale homogenously 
or almost homogenously, respectively, with degree $D$.

\subsection{Stückelberg--Petermann renormalization group and Main Theorem}

The Main Theorem is essentially the following statement:
Let $\nS$ and $\hat\nS$ be two solutions of the axioms for the $T$-product.
Then there exists a renormalization of the interaction
$F\mapsto Z(F)$ such that
\begin{equation}
 \hat\nS(F)=\nS\bigl(Z(F)\bigr)\quad \forall F\in \sF_\loc \ .
\end{equation}
The definition of the St\"uckelberg--Petermann renormalization group (RG)
is such that the set of all maps $Z\: \sF_\loc\longto \sF_\loc$
appearing in this relation, when $(\nS,\hat\nS)$ runs through all
admissible pairs of $S$-matrices, is precisely the St\"uckelberg--Petermann RG.
Consequently, the definition of the St\"uckelberg--Petermann RG depends on
the set of renormalization conditions for the $T$-product.
For brevity, we use here only Field Independence and Translation covariance.

\begin{defn}\label{df:SP-RG}
The \emph{St\"uckelberg--Petermann RG} is the set $\sR$ of all maps\\
$Z\: \sF_\loc\pw{\hbar,\ka} \longto \sF_\loc\pw{\hbar,\ka}$
satisfying the following properties:
\begin{itemize}
\item[(1)] \emph{Taylor series}: $Z(F)$ is a formal Taylor series w.r.t. $F=0$, that is,
$$
Z(F) = \sum_{n=0}^\infty \frac{1}{n!}\, Z^{(n)}(F^{\ox n})\ ,
\word{where}  Z^{(n)}:=Z^{(n)}(0)\,:\,\sF_\loc\pw{\hbar,\ka}^{\ox n}\longto\sF_\loc\pw{\hbar,\ka}
$$
is the $n$th derivative of $Z(F)$ at $F=0$ (i.e., $Z^{(n)}$ is linear,
symmetrical in all factors and can be computed by
$Z^{(n)}(F^{\ox n})= \frac{d^n}{d\la^n}\bigr|_{\la=0} Z(\la F)$).

\item[(2)] \emph{Lowest orders}:
$Z^{(0)}\equiv Z(0) = 0$, \ $Z^{(1)} = \Id\ $.

\item[(3)] \emph{Locality, Translation covariance}:
For all monomials $A_1,\ldots$
$\ldots,A_n\in\sP$ the VEV
\be\label{eq:z-def}
z^{(n)}(A_1,\ldots,A_n)(x_1-x_n,\ldots):=\om_0\Bigl(
Z^{(n)}\bigl(A_1(x_1),\ldots, A_n(x_n)\bigr)\Bigr)
\ee
depends only on the relative coordinates and is of the form
\be\label{eq:Z-local}
z^{(n)}(A_1,\ldots,A_n)(x_1-x_n,\ldots)
=\gS_n\sum_{a\in\bN^{d(n-1)}}C^a(A_1,\dots,A_n)\,
\del^a\dl(x_1 - x_n,\dots,x_{n-1} - x_n)\,,\nonumber
\ee
with constant coefficients $ C^a(A_1,\dots,A_n)$ and $\gS_n$ denotes symmetrization in $(A_1,x_1),\ldots,(A_n,x_n)$.

\item[(4)] \emph{Field independence}:
$\dl Z/\dl\vf = 0$.
\end{itemize}
\end{defn}

The distributions $Z^{(n)}\bigl(A_1(x_1),\ldots, A_n(x_n))\in
\sD'(M^n,\sF_\loc)\ ,\,\,A_1,\ldots,A_n\in\sP$ appearing in \eqref{eq:z-def} are defined 
in terms of $Z^{(n)}:\sF_\loc^{\ox n}\to\sF_\loc$ analogously
to $T_n\bigl(A_1(x_1),\ldots, A_n(x_n))$ (see \eqref{eq:def-AWI}), hence, they also satisfy the AWI.

Similarly to the $T$-product,
the Field independence property (4) for $Z$ is equivalent to the validity of the
(causal) Wick expansion \eqref{eq:causal-Wick-expan} for $Z^{(n)},\,\,\forall n\geq 2$.

In the framework of causal perturbation theory, a first version of the
Main theorem was given by Popineau and
Stora~\cite{PopineauS16}; we present here a more elaborated version.

\begin{thm}[Main Theorem]\label{th:main-thm-renorm}
\begin{enumerate}
\item[\textup{(a)}]
Given two $S$-matrices $\nS$ and $\wh \nS$ both fulfilling the axioms,
there exists a formal Taylor series  $Z \: \sF_\loc\pw{\hbar,\ka} \longto \sF_\loc\pw{\hbar,\ka}$ 
with expansion point $0$ $($i.e.,
$Z$ satisfies the property $(1)$ of Definition $\ref{df:SP-RG})$, which is
uniquely determined by
\be\label{eq:main-theorem}
\wh\nS = \nS \circ Z\ .
\ee
This $Z$ is an element of the St\"uckelberg--Petermann RG~$\sR$.

\item[\textup{(b)}]
Conversely, given an $S$-matrix $\nS$ fulfilling the axioms
for the $T$-product and an arbitrary $Z \in \sR$,
the composition $\wh\nS := \nS \circ Z$ also satisfies
these axioms.
\end{enumerate}
\end{thm}

A corollary of this Theorem states that $(\sR,\circ)$ is indeed a group.

\subsection{Interacting fields and the algebraic adiabatic limit}\label{sec:ad-lim}
Interacting fields are obtained from the $S$-matrix by Bogoliubov's definition \cite{BogoliubovS59}:  for $F\in\sF_\loc$
the formal power series
\be\label{eq:Bogo}
F_S:= \frac{\hbar}{i}\, \ddto{\la} \nS(S)^{\star-1} \star \nS(S + \la F)\in\sF\pw{\hbar,\ka}
\ee
is the \emph{interacting field} belonging to the free field $F$ (i.e., $F_S\vert_{\ka=0}=F$) and to the interaction $S$.
By using only the basic axioms for $T$, one proves that the so defined interacting fields satisfy Causality, that is,
\be\label{eq:Causality-intfield}
F_{S+G}=F_S\word{if} \supp G\cap (\supp F+\ovl V_-)=\emptyset,
\ee
and the Glaser-Lehmann-Zimmermann (GLZ) relation (which plays an important role in Steinmann's inductive construction of the 
perturbative interacting fields \cite{Steinmann71}):%
\footnote{Actually, the proof of the GLZ relation uses \emph{only} the
definition \eqref{eq:Bogo} and Linearity (axiom (1)) and Symmetry (axiom (3)).  The GLZ relation can be interpreted as an integrability condition
for the ``vector fields'' $S \to F_S$, which ensures the
existence of a ``potential'' $\nS(S)$ from which $F_S$ can be recovered
by~\eqref{eq:Bogo}.}
\be
\frac{1}{i\hbar} \bigl[G_S\,,\,F_S\bigr]_\star =\frac{d}{d\la}\Big\vert_{\la=0}\Bigl(F_{S+\la G}-G_{S+\la F}\Bigr)
\ee
(where $[\,\cdot\,,\,\cdot\,]_\star$ denotes the commutator w.r.t.~the star product). Combining these two properties we 
get spacelike commutativity:
$$
[G_{S}, \,F_{S}]_\star = 0 \word{if} (x - y)^2 < 0
\word{for all} (x,y) \in \supp G \x \supp F.
$$
%The GLZ relation  provides a splitting of the commutator
%$[A_S(x),B_s(y)]_\star$ (where $A,B\in\sP_\bal$)
%into a ``advanced'' (i.e., $(x-y)\in\ovl{V}_-$)
%and a ``retarded part'' ($(x-y)\in\ovl{V}_+$).
The validity of the renormalization conditions for $T$ implies corresponding properties for the interacting fields,
e.g., unitarity (axiom (6)) implies $(F_S)^*=(F^*)_{S^*}$ and, as the name says, axiom (8) implies
$$
(\square+m^2)\vf(x)_S=(\square+m^2)\vf(x)+\Bigl(\frac{\dl S}{\dl\vf(x)}\Bigr)_S\ .
$$
In addition, axiom 11 ($\hbar$-dependence) implies that, for $S,\,F\sim \hbar^0$, the interacting field $F_S$ is a formal power series in 
$\hbar$; hence, its limit $\hbar\to 0$ exists and gives the pertinent (perturbative) \emph{classical} interacting field -- however, the 
limit $\lim_{\hbar\to 0}\nS(S)$ does not exist.

\paragraph{Algebraic adiabatic limit.} To obtain scattering amplitudes contributing to 
inclusive cross sections, one has to perform the (weak)
adiabatic limit $\lim_{g\to 1}\om_0\bigl(F\star\nS(S(g))\star G\bigr)$ for appropriate $G,F\in\sF$ describing the
in- and out-state, respectively. In contrast, from the interacting fields one can extract observable quantities without 
performing this limit; to wit, the \emph{local, algebraic} structure of these fields does not 
depend on the adiabatic switching of the interaction. 

To explain this, let $\sO$ be an open double cone (i.e.,
$\sO = (x + V_+) \cap (y + V_-)$ for some pair $(x,y)\in M^2$ fulfilling $y \in (x + V_+)$)
and let $\sF_\loc(\sO):=\set{F\in\sF_\loc\,\big\vert\,\supp F\subset \sO}$.
We introduce the algebra of interacting fields localized in $\sO$:
\be\label{eq:A(O)}
\sA_{L_\intr}(\sO):= \bigveestar \set{F_{S(g)}\,\big\vert\, F \in \sF_\loc(\sO)}\word{with}
g \in \sG(\sO) := \set{g \in \sD(\bM,\bR)\,\big\vert\, g\vert_{\ovl\sO}=1},
\ee
where $\bigveestar$ means the algebra, under the $\star$-product,
generated by members of the indicated set and $S(g)$ is obtained from $L_\intr$ by \eqref{eq:S(g)}.
A main problem is that $F_{S(g)}$ depends on the
restriction of~$g$ to $\sO + \ovl V_-$ (by causality, see \eqref{eq:Causality-intfield}); but the algebra
$\sA_{L_\intr}(\sO)$ should rather be independent of~$g$. This is indeed the case \cite{BF00}.

\begin{thm}\label{th:local-algebra}
As an abstract algebra, $\sA_{L_\intr}(\sO)$ \eqref{eq:A(O)} is independent of the choice of $g \in \sG(\sO)$. Concretely, for any
$g_1, g_2 \in \sG(\sO)$, there is a unitary%
\footnote{To say that $U$ is \emph{unitary} means that
$U^* \star U = U \star U^* = (1,0,0,\dots)$ in $\sF\pw{\ka}$.}
element $U_{g_1,g_2} \in \sF\pw{\ka}$ such that
\begin{equation}
U_{g_1,g_2} \star F_{S(g_1)} \star (U_{g_1,g_2})^{\star-1}
= F_{S(g_2)}\,,\quad\text{for all}\quad F \in \sF_\loc(\sO)\,.
\label{eq:ad-lim-basis} % (3.ao)
\end{equation}
\end{thm}

\noindent The proof uses only Causality and Unitarity of the $S$-matrix, i.e., \eqref{eq:causality-S} and \eqref{eq:unitarity}.

\section*{Acknowledgements}
Hints given by Romeo Brunetti, Jan Derezi\'nski, Klaus Fredenhagen, Christian Ga\ss  $\,$ and Kasia Rejzner have been used to improve the manuscript.

\end{document}